\documentclass[twocolumn]{aastex61}
\usepackage{graphicx}	% Including figure files
\usepackage{amsmath}	% Advanced maths commands
\usepackage{amssymb}	% Extra maths symbols
\usepackage{slashed}
\usepackage{bm}

\newcommand{\nver}{\hat{\mathbf{n}}}
\newcommand{\advalue}{\ensuremath{1.00 \pm 0.21}}
\newcommand{\dgvalue}{\ensuremath{1.03 \pm 0.19}}
\newcommand{\bkg}{\ensuremath{1.33\pm 0.35}}
\newcommand{\bgg}{\ensuremath{1.13\pm 0.02}}
\newcommand{\sqdeg}{\ensuremath{\rm deg^2}}

\received{}
\revised{}
\accepted{}
\submitjournal{ApJ}
\shorttitle{$D_G$ with 2MPZ and Planck lensing}
\shortauthors{F. Bianchini \& C. L. Reichardt}
\begin{document}

\title{Constraining gravity at large scales with the 2MASS Photometric Redshift catalogue and \textit{Planck} lensing}

\correspondingauthor{Federico Bianchini}
\email{fbianchini@unimelb.edu.au}

\author[0000-0003-4847-3483]{Federico Bianchini}
\affil{School of Physics, University of Melbourne, Parkville, VIC 3010, Australia}

\author[0000-0003-2226-9169]{Christian L. Reichardt}
\affiliation{School of Physics, University of Melbourne, Parkville, VIC 3010, Australia}



\begin{abstract}
We present a new measurement of structure growth at $z \simeq 0.08$ obtained by correlating  the cosmic microwave background (CMB) lensing potential map from the \textit{Planck} satellite with the angular distribution of the 2MASS Photometric Redshift galaxies. 
After testing for, and finding  no evidence for systematic effects,  
we calculate the angular auto- and cross-power spectra.
We combine these spectra to estimate the amplitude of structure growth using the bias-independent $D_G$ estimator introduced by \citet{Giannantonio2016}. 
We find that the relative amplitude of $D_G$ with respect to the predictions based on \textit{Planck} cosmology is $A_D(z=0.08) = \advalue$, fully consistent with the expectations for the standard cosmological model.
Considering statistical errors only, we forecast that a joint analysis between an LSST-like photometric galaxy sample and lensing maps from upcoming ground-based CMB surveys like the Simons Observatory and CMB-S4 can yield sub-percent constraints on the growth history and differentiate between different models of cosmic acceleration.   
\end{abstract}

\keywords{Cosmic background radiation -- large-scale structure of Universe -- gravitational lensing: weak}



\section{Introduction} \label{sec:intro}
Almost twenty years after its discovery, the accelerated expansion of the universe \citep{Riess1998,Perlmutter1999} remains one of the most pressing  questions in physics. 
Given its significance, a variety of cosmological probes are being deployed to understand the origin of this acceleration. 
So far the observational evidence is consistent with the acceleration that is sourced by a cosmological constant $\Lambda$ \citep{deHaan2016,planck_params_2015,Jones2017}, which is also perhaps the most economical solution.
However, this explanation has its shortcomings \citep{Weinberg1989,Martin2012}. 
There are two main alternative frameworks, namely \textit{dark energy} or \textit{modified gravity}. 
Dark energy models add a dynamical degree of freedom to the stress-energy tensor that begins to dominate the cosmic energy budget in recent times \citep{Ratra1988}.  
Modified gravity models instead change general relativity on cosmological distances \citep{Silvestri2009}. 
Detecting any deviation from the predictions of $\Lambda$CDM is the first step toward distinguishing between the dark energy and modified gravity paradigms.

The signatures of these two effects are degenerate at the cosmic expansion history level, probed, for example, by supernovae and baryonic acoustic oscillations. 
The expansion history must be combined with growth of structure measurements, such as redshift space distortions (RSD), weak gravitational lensing, and galaxy cluster counts to distinguish between different models of cosmic acceleration. 

Combinations of different probes are also emerging as  robust and promising tools for conducting precision tests of the standard cosmological model ($\Lambda$CDM). 
Cosmological observables can be either jointly analyzed in a Bayesian framework \citep{Doux2017,Nicola2017,DES2017} or combined into a single statistics.

We follow the latter approach in this paper. Specifically, we measure the $D_G$ statistic, introduced by \citet{Giannantonio2016} in the context of photometric redshift surveys, to constrain the linear growth factor $D(z)$. This estimator combines CMB lensing and galaxy clustering measurements in such a way that it is relatively insensitive to galaxy bias on linear scales. While lensing probes the cumulative matter distribution along the line of sight (LOS), galaxy surveys provide a biased sampling of the dark matter field. Then, a joint measurement of lensing and clustering helps in breaking the degeneracy between growth and bias. 
$D_G$ was first measured by \citet{Giannantonio2016} using the Dark Energy Survey (DES) galaxy sample between $0.2 \le z \le 1.2$, and the CMB lensing maps from \textit{Planck} and the South Pole Telescope (SPT). When compared with the fiducial growth history fixed by \textit{Planck} observations, they found a value which is $1.7\sigma$ lower than expected for a $\Lambda$CDM cosmology.

Another estimator similar in spirit to $D_G$, is the $E_G$ statistic proposed by \citet{Zhang2007}. $E_G$ is defined as the ratio between the Laplacian of the Newtonian potentials to the peculiar velocity divergence. 
Because it probes the relative dynamics of relativistic and massive particles, $E_G$ acts as a consistency test of the laws of gravity. 
\citet{Reyes2010} have made the first measurement of the  $E_G$ statistic  by combining the cross-correlation between lensing and foreground galaxies, with RSD and the clustering amplitude of the lenses. 
$E_G$ has since been  applied  in the context of spectroscopic surveys and galaxy lensing datasets \citep{Reyes2010,Blake2016,Amon2017}, and more recently has been extended to CMB lensing \citep{Pullen2016}. 

Spectroscopic surveys represent the main avenue to probe the 3D matter distribution; however they are more costly in terms of time and resources than photometric surveys which are usually utilized to conduct 2D (angular) studies. As a consequence, spectroscopic surveys are shallower and/or narrower than their photometric counterparts, resulting in lower number density. Current and upcoming imaging surveys will deliver galaxy samples with a radial resolution accurate enough to allow for tomographic 2D analyses of the clustering and cross-correlation with external probes, recovering most of the 3D information \citep{Asorey2012}.
A cosmological quantity traditionally measured through the statistical analysis of the anisotropic galaxy correlation function in spectroscopic redshift survey is the linear growth rate, defined as the logarithmic derivative of the growth factor with respect to the cosmic scale factor $a$, $f=\text{d}\ln D/\text{d}\ln a$. Of course, this quantity is tightly related to the growth factor, as they contain the same information. Growth rate measurements at $z\sim 0$ using RSD have currently reached an accuracy of about $\sim 10\%$, such as from 2dF Galaxy Redshift Survey \citep[2dFGRS]{Beutler2012}  or SDSS Main Galaxy Sample \citep{Howlett2015}, although the theoretical modeling of the RSD signal presents a number of challenges. Therefore, it is crucial to complement RSD measurements with independent analyses like the one presented in this paper.

The main focus of this work is to measure $D_G$ at $z \sim 0.08$ by combining the CMB lensing map reconstructed by the \textit{Planck} team \citep{planck_lens_2015} with the spatial distribution of the 2MASS Photometric Redshift galaxies \citep[2MPZ]{2MPZ}. This analysis extends the measurements of the $D_G$ statistics to lower redshifts, where most of the departures from GR are expected, and over a significantly larger sky area -- 26,000\,\sqdeg{} -- than probed before \citep{Giannantonio2016}.

The remainder of this work is structured as follows. In Sec.~\ref{sec:datasets} we describe the datasets exploited in the analysis, while the theoretical background and methodology are reviewed in Sec.~\ref{sec:theorynmethods}. Systematic checks are discussed in Sec.~\ref{sec:systematics}, while we present the results and  forecasts for future surveys in Sec.~\ref{sec:results}. Finally, we summarize our findings in Sec.~\ref{sec:conclusions}.

Throughout the paper, unless stated otherwise, we assume a fiducial flat $\Lambda$CDM described by the \texttt{Planck2013+WP+highL+BAO} \citep{planck_params_2013} parameters, $\{\Omega_{\rm b}h^2,\Omega_{\rm c}h^2, h, n_s, A_s\}=\{0.0222,0.119,\\
0.678,0.961,2.21\times 10^{-9}\}$, corresponding to $\Omega_{\rm m} = 0.307$ and $\sigma_8 = 0.829$.

\section{Data sets}
\label{sec:datasets}
\subsection{2MPZ}
The 2MASS Photometric Redshift catalog\footnote{\url{http://ssa.roe.ac.uk/TWOMPZ.html}} \citep{2MPZ} is an almost all-sky galaxy sample that includes photometric information for approximately 935,000 sources. The catalog has been built by cross-matching the near-IR 2MASS Extended Source Catalog \citep[XSC]{XSC} with optical SuperCOSMOS \citep[SCOS]{SCOS} and mid-IR Wide-field Infrared Survey Explorer \citep[WISE]{WISE} data. This results in a flux-limited catalog to $K_S \le 13.9$ mag (Vega), roughly corresponding to the 2MASS XSC full-sky completeness limit. The multi-wavelength information allows to estimate accurate photo-$z$s for the sources by employing neural network algorithms trained on subsamples drawn from spectroscopic redshift surveys overlapping with 2MASS. Inferred photo-$z$s are virtually unbiased ($\delta z \simeq 0$) and their random errors are $\sim 15\%$ (RMS normalized scatter of $\sigma_z \simeq 0.015$). 2MPZ sources lie in the redshift range of $z \lesssim 0.4$ with the median being $z_{\rm med}\simeq 0.08$ and 95\% of the sources below $z\lesssim 0.17$. We define our baseline sample by retaining all of the galaxies in the photometric redshift range $0 \le z \le 0.24$, but do not apply any magnitude cut. Finally, we construct an overdensity map $\delta_g(\nver)= n(\nver)/\bar{n}-1$,  where $n(\nver)$ is the number of objects in a given pixel and $\bar{n}$ is the mean number of objects per pixel in the unmasked area, in the \texttt{HEALPix}\footnote{\url{http://healpix.jpl.nasa.gov}} \citep{Gorski2005a} format with a resolution parameter $N_{\rm side}=256$ (approximately $13\arcmin.7$ pixel size). 

\subsection{\textit{Planck} CMB lensing}
We use the publicly available\footnote{\url{http://pla.esac.esa.int/pla/}} CMB lensing convergence map reconstructed by the \textit{Planck} team \citep{Planck_lensing_2015}. The lensing convergence map has been extracted by applying the quadratic lensing estimator developed by \citet{Okamoto2003} to \texttt{SMICA} foreground-cleaned CMB temperature ($T$) and polarization ($P$) maps. Different quadratic combinations of the $T$ and $P$ maps are then combined to form a minimum-variance (MV) estimate of the lensing convergence $\kappa$ bandpass filtered between $8 \le \ell \le 2048$.
The MV lensing reconstruction is provided in the \texttt{HEALPix} format at a resolution of $N_{\rm side}=2048$, corresponding to $1\arcmin.7$ pixel size. 

\subsection{Masks}
Even though the 2MPZ catalog and \textit{Planck} have almost full-sky coverage, there are regions unsuitable for cosmological studies due to observational effects. 
The main obstacle is the obstruction of view by our own Galaxy, creating the so-called Zone of Avoidance. On top of this, we have to exclude regions contaminated by Galactic foregrounds, such as dust, stars, bad seeing, as well as areas with incomplete coverage.

We construct a 2MPZ fiducial mask over which we carry out our cosmological analysis following \cite{Alonso2015}. Briefly, we assume the reddening $E(B-V)$ map from \cite{Schlegel1998} to trace the Galactic extinction in the $K$-band as $A_K = 0.367 \,E(B-V)$, and consider the star density $n_{\rm star}$ at each 2MPZ galaxy position. We then create \texttt{HEALPix} maps of $A_K$ and $n_{\rm star}$ at a resolution $N_{\rm side}=64$ and discard all pixels for which $A_K>0.06$ and $\log_{10}n_{\rm star} > 3.5$. This eliminates regions near the Galactic plane and the Magellanic Clouds, and covers $f_{\rm sky} = 0.69$. 

 We also have a mask for the \textit{Planck} CMB lensing dataset. The \textit{Planck} mask is the combination of (i) a 70\% Galactic mask, (ii) a point source mask at 143 and 217 GHz, (iii) a mask that removes Sunyaev-Zel'dovich (SZ) clusters with signal-to-noise $S/N>5$ in the \textit{Planck} SZ catalog PSZ1, (iv) as well as the \texttt{SMICA} $T$ and $P$ confidence masks. The sky fraction left for the lensing reconstruction is approximately $f_{\rm sky}=0.67$.

We then multiply the 2MPZ and \textit{Planck} masks to construct our fiducial mask.
 With this fiducial mask,  the main galaxy sample has 639,673 sources over $f_{\rm sky}=0.62$. 
 This corresponds to a galaxy number density of $\bar{n} = 8.2\times 10^{5}$ sr$^{-1}$, equivalent to 24.5 deg$^{-2}$ or $1.3$ pix$^{-1}$.
 
\section{Theory and methods}
\label{sec:theorynmethods}
In this section, we briefly review how a combination of CMB lensing and galaxy clustering measurements can constrain the cosmic growth history. 
We then outline the analysis methods.

\subsection{Theoretical background}
\label{sec:theory}
\begin{figure}
	\includegraphics[width=\columnwidth]{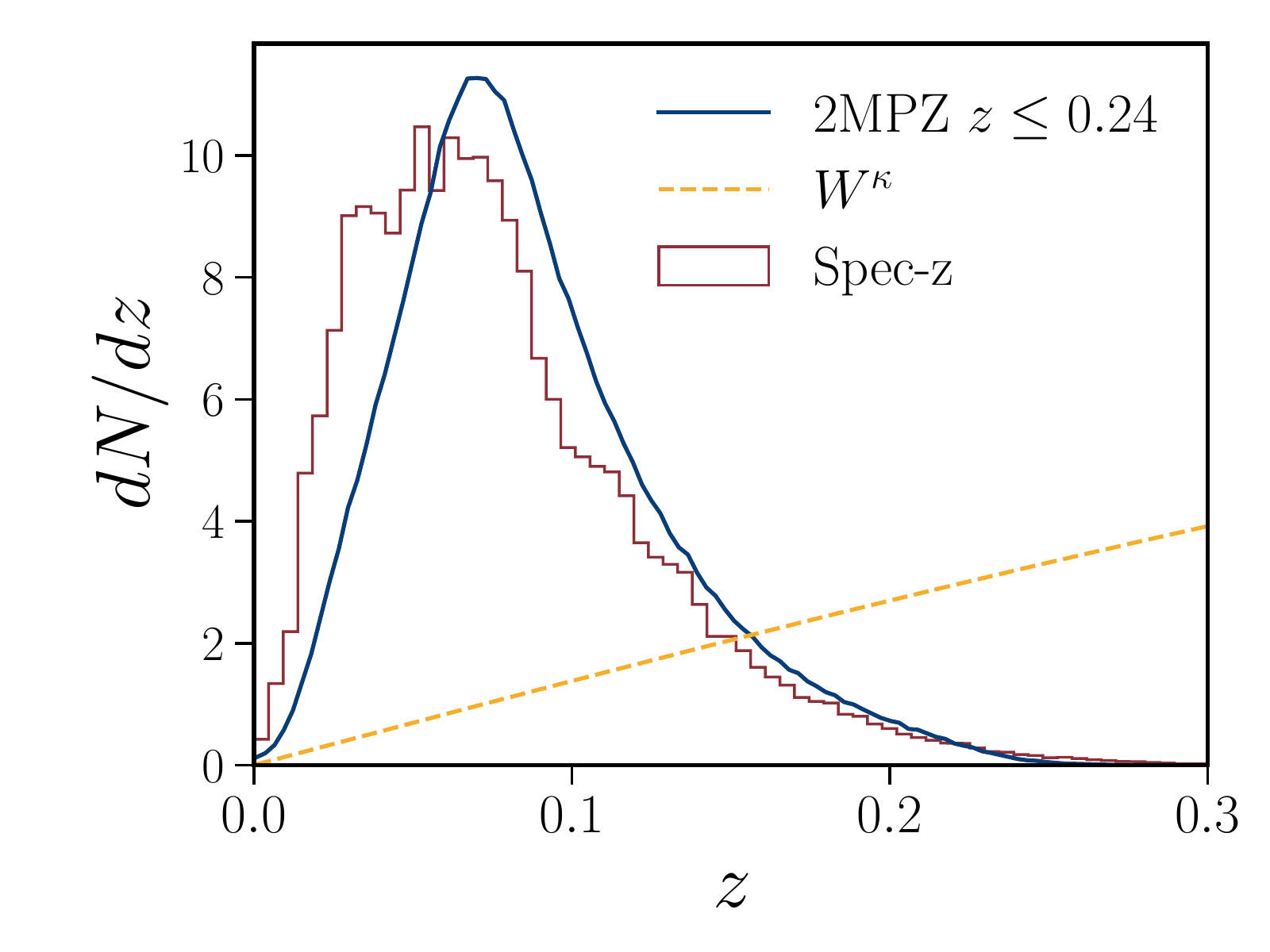}
    \caption{
    Redshift overlap of the 2MPZ galaxy catalog with the CMB lensing kernel is small and increasing with redshift.
The solid blue line shows the redshift distribution of the 2MPZ galaxy catalog selected at $z_{\rm ph}\le 0.24$ while the dashed orange line shows the CMB lensing kernel.     
The red histogram shows the redshift distribution of 2MPZ galaxies for which spectroscopic information is available ($\approx 1/3$ of full sample). All information is in arbitrary units.}
    \label{fig:2MPZ_dNdz}
\end{figure}

The basic idea behind combining CMB lensing and galaxy clustering measurements
is that the two observables respond to the underlying dark matter field in complementary ways. 
Whereas lensing measurements are sensitive to the integrated matter distribution along the LOS, galaxy surveys provide a biased sparse sampling of the dark matter field. 
More quantitatively under general relativity, the CMB lensing convergence $\kappa$ and galaxy overdensity $\delta_g$ fields can be written as LOS projections of the 3D dark matter density contrast $\delta$:
\begin{equation}
X(\nver) = \int_0^{\infty} dz\, W^X(z)\delta(\chi(z)\nver,z).
\end{equation}
In the above equation, $X=\{\kappa,g\}$. 
The kernels $W^X(z)$ encode each observable's response to the underlying dark matter distribution:
\begin{equation}
\label{eqn:wk}
W^{\kappa}(z) = \frac{3\Omega_m}{2c}\frac{H_0^2}{H(z)}(1+z)\chi(z)\frac{\chi_*-\chi(z)}{\chi_*},
\end{equation}
\begin{equation}
W^{g}(z) = b(z)\frac{dN}{dz}.
\label{eqn:wg}
\end{equation}
Here, $H(z)$ is the Hubble factor at redshift $z$, $\chi(z)$ and $\chi_*$ are the comoving distances to redshift $z$ and to the last scattering surface. $\Omega_m$ and $H_0$ are the present-day values of matter
density and Hubble parameter, respectively. In Eq.~\ref{eqn:wg}, we assumed a linear, local, and deterministic galaxy bias $b(z)$ to relate the galaxy overdensity $\delta_g$ to the matter overdensity $\delta$ \citep{Fry1993}, while the galaxy sample unit-normalized redshift distribution is denoted as $dN/dz$. 
Since we are selecting galaxies based on photo-$z$'s, we must also account for the photo-$z$ uncertainties. 
We do this by convolving the sample's photometric redshift distribution with the catalog's photo-$z$ error function \citep{Sheth2010}:
\begin{equation}
\frac{dN}{dz} = \int_0^{\infty} dz \, W(z_{\rm ph})\frac{dN}{dz}(z_{\rm ph})p(z|z_{\rm ph}).
\end{equation}
In this equation $W(z_{\rm ph})$ defines the redshift bin -- usually a top-hat function in photo-$z$ space -- while the photo-$z$ error function is modeled as a Gaussian of redshift-dependent width, $p(z|z_{\rm ph}) \sim \mathcal{G}(0,\sigma_{z}(1+z))$, where $\sigma_z = 0.015$ \citep{2MPZ,Balaguera2017}. The resulting redshift distribution for our  galaxy sample is shown as the solid blue line in Fig.~\ref{fig:2MPZ_dNdz}.
The knowledge of the galaxy redshift distribution is a key ingredient needed to translate the observed power spectra into a growth factor estimate.
We quantify the impact of the assumed redshift distribution on the results of the analysis in Sec.~\ref{sec:robustness}.

For scales smaller than $\ell \gtrsim 10$, we can adopt the so-called Limber approximation \citep{limber} and evaluate the theoretical angular auto- and cross-power spectra as:
\begin{equation}
\label{eqn:limber}
C_{\ell}^{XY} =   \int_0^{\infty} \frac{dz}{c} \frac{H(z)}{\chi^2(z)} W^{X}(z)W^{Y}(z)P\left(k=\frac{\ell+\tfrac{1}{2}}{\chi(z)},z\right).
\end{equation}
Eq.~\ref{eqn:limber} can be thought of as a weighted integral of the matter power spectrum $P(k,z)=P(k,0)D^2(z)$, where $D(z)$ is the linear growth function normalized to unity at $z=0$:
\begin{equation}
D(z) = \exp\left\{{-\int_0^z \frac{\left[\Omega_{\rm m}(z')\right]^\gamma}{1+z'}}dz'\right\},
\end{equation}
and $\gamma \approx 0.55$ is the growth index in the case of general relativity \citep{Linder2005}.
We compute the non-linear $P(k,z)$ using the \texttt{CAMB}\footnote{\url{https://camb.info/}} code with the \texttt{Halofit} prescription \citep{camb,halofit}.\footnote{This implies that we can factorize the non-linear matter power spectrum as $P_{\rm NL}(k,z)=D^2(z)P_{\rm NL}(k,0)$. We checked that this assumption holds to more than	$3\%$ accuracy over the scales and redshifts of interest.}
As shown in \citet{Balaguera2017}, the impact of RSD and the Limber approximation is $\lesssim 5\%$ and confined to scales $\ell \lesssim 10$. 
Given that this level of theoretical uncertainty is smaller than the statistical errors  and that we limit the analysis to $\ell > 10$, we ignore these effects here. In Eq.~\ref{eqn:wg} we also neglect the effect of the lensing magnification bias (see \citet{Bianchini2016} for the expression including these effects).

Examining Eq.~\ref{eqn:limber}, one notices that the auto-power spectrum scales as $C_{\ell}^{gg} \propto b^2(z)D^2(z)$ while the cross-spectrum scales as $C_{\ell}^{\kappa g} \propto b(z)D^2(z)$. 
Thus an appropriate combination of the two can eliminate the bias and  break the degeneracy between galaxy bias and cosmic growth.
\citet{Giannantonio2016} devised a bias-independent estimator for photo-$z$ surveys to recover the cosmic growth information:
\begin{equation}
\label{eqn:dg}
\hat{D}_G = \left \langle \frac{\hat{C}_{\ell}^{\kappa g}}{\slashed{C}_{\ell}^{\kappa g}}\sqrt[]{\frac{\slashed{C}_{\ell}^{gg}}{\hat{C}_{\ell}^{gg}}} \right \rangle_{\ell}.
\end{equation}
Here, the hat represents measured power spectra while slashed quantities denote theoretical power spectra calculated by removing the growth function from the Limber integration, i.e. the matter power spectrum in Eq.~\ref{eqn:limber} is evaluated at $z=0$. 
We emphasize that Eq.~\ref{eqn:dg} is averaged over the range of multipoles included in the analysis.

In the limit of no  galaxy bias evolution over a bin (narrow redshift bins), the $D_G$ estimator has the advantage of being bias-independent: the true bias  shows up in both denominator and numerator and thus cancels out, as does the assumed bias. Its expectation value is $\langle D_G \rangle = D$ on linear scales, although one might be concerned about the impact of non-linearities. We test for the dependence of the growth factor constraint on the choice of angular scale cuts below. Note that the $D_G$ estimator will scale as  $\sigma_8\Omega_m H_0^2$ due to its dependence on the matter power spectrum and CMB lensing kernel.

\subsection{Methods}
\label{sec:methods}
In this work, we measure the $D_G$ statistic in the harmonic domain by combining the observed $\hat{C}_{\ell}^{\kappa g}$ and $\hat{C}_{\ell}^{gg}$ spectra. 
We work with maps at an \texttt{HEALPix} resolution of $N_{\rm side}=256$ and convert between different resolutions using the \texttt{HEALPix} built-in \texttt{ud\_grade} routine. Power spectra are extracted using a pseudo-$C_{\ell}$ method based on \texttt{MASTER} algorithm \citep{Hivon2002} that deconvolves for the mask induced mode-coupling and pixelization effects. 

Operating with cross-power spectra as for $\hat{C}_{\ell}^{\kappa g}$ has a number of advantages. 
A cross-spectrum is free of noise bias and it is less prone to systematics as the systematics and noise rarely correlate between different experiments and observables.

The analysis also uses galaxy-galaxy auto-spectrum $\hat{C}_{\ell}^{gg}$ which has to be noise subtracted. Here, we do not debias for the shot-noise term, $N_{\ell}^{gg}=1/\bar{n}$, but rely on a jackknifing approach instead (see, for example, \citet{Ando2018}). We randomly split the galaxy catalog in two and create two galaxy overdensity maps $\delta_g^1$ and $\delta_g^2$. From these, we form a pair of half-sum and half-difference maps, $\delta_g^{\pm} = (\delta_g^1\pm \delta_g^2)/2$. The former map will contain both signal and noise, while the latter will be noise-only. We then extract their auto-power spectra and evaluate the total galaxy auto-power spectrum as $\hat{C}_{\ell}^{gg}= \hat{C}_{\ell}^{++}-\hat{C}_{\ell}^{--}$. 

We estimate both angular power spectra in linearly spaced band powers of width $\Delta\ell=10$ between $10 \le \ell \le 250$, where the lower limit is imposed by the filtering applied to the \textit{Planck} lensing map. 
The maximum multipole is not limited by the data (as long as $\ell_{\rm max} \lesssim 2N_{\rm nside}$). 
Instead, the choice of $\ell_{\rm max}$ is motivated by the desire to avoid strongly non-linear scales. 
In order to reduce potential contaminations from non-linearities, in our baseline analysis we set $\ell_{\rm max}$ to the angular scale subtended by the density modes that are entering the non-linear regime at $z\simeq 0.08$, i.e. $\Delta^2(k_{\rm NL}) = k^3_{\rm NL}P^{\rm lin}(k_{\rm NL})/(2\pi^2 )\approx 1$. Therefore, we set $\ell_{\rm max}=70$ and explore below the robustness of the results against different choices of $\ell_{\rm max}$. 
We have also checked that adopting a finer bin width of $\Delta\ell=5$ has a negligible impact on the results of the analysis.

Assuming that both fields behave as Gaussian random distributed variables on the scales of interest, we evaluate the error bars as
\begin{equation}
\label{eqn:errors}
\left( \Delta\hat{C}_{L}^{XY} \right)^2 = \frac{1}{(2L+1)\Delta\ell f_{\rm sky}} \left[(\hat{C}^{XY}_{L})^2 + \hat{C}^{XX}_{L}\hat{C}^{YY}_{L} \right],
\end{equation}
where $\Delta\ell$ is the bin width of a band power centered at $L$ and $\hat{C}_L^{XX} (\hat{C}_L^{XY}$) is the auto- (cross-)spectrum comprehensive of noise bias. By setting $X=Y$ in Eq.~\ref{eqn:errors}, one finds the expression for the auto-power spectra uncertainties. The validity of this assumption has been tested by \citet{Balaguera2017}, who have compared the Gaussian error bars with uncertainties estimated trough jackknife re-sampling and galaxy mocks methods. 

As we mentioned in Sec.~\ref{sec:theory}, the $D_G$ estimator involves averaging over some angular scales. To this end, we make use of the uncertainties information given by Eq.~\ref{eqn:errors} and apply an inverse-variance weighting when averaging across multipoles $\ell$'s. The details concerning the weighting scheme can be found in App.~\ref{sec:weights}.

For the survey specifications discussed in Sec.~\ref{sec:datasets}, and assuming a galaxy bias $b\simeq 1.2$ \citep{Alonso2015}, we forecast an overall signal-to-noise ($S/N$) in the multipole range $\ell \in [10,70]$ of $\approx 4.3$ and $\approx 37$ for the cross- and auto-correlation respectively. 
As will be seen in the next section, this forecast is consistent with the real $D_G$ measurement uncertainty. 

\section{Systematics checks}
\label{sec:systematics}
Before presenting the results, we summarise here the null tests done to assess the robustness of the analysis against systematic effects. Here we focus on testing the cross-power spectrum since \citet{Balaguera2017} have thoroughly searched for systematic effects affecting the 2MPZ catalogue, such as varying depth and dust extinction, and found it suitable for clustering analyses.
\begin{figure}
	\includegraphics[width=\columnwidth]{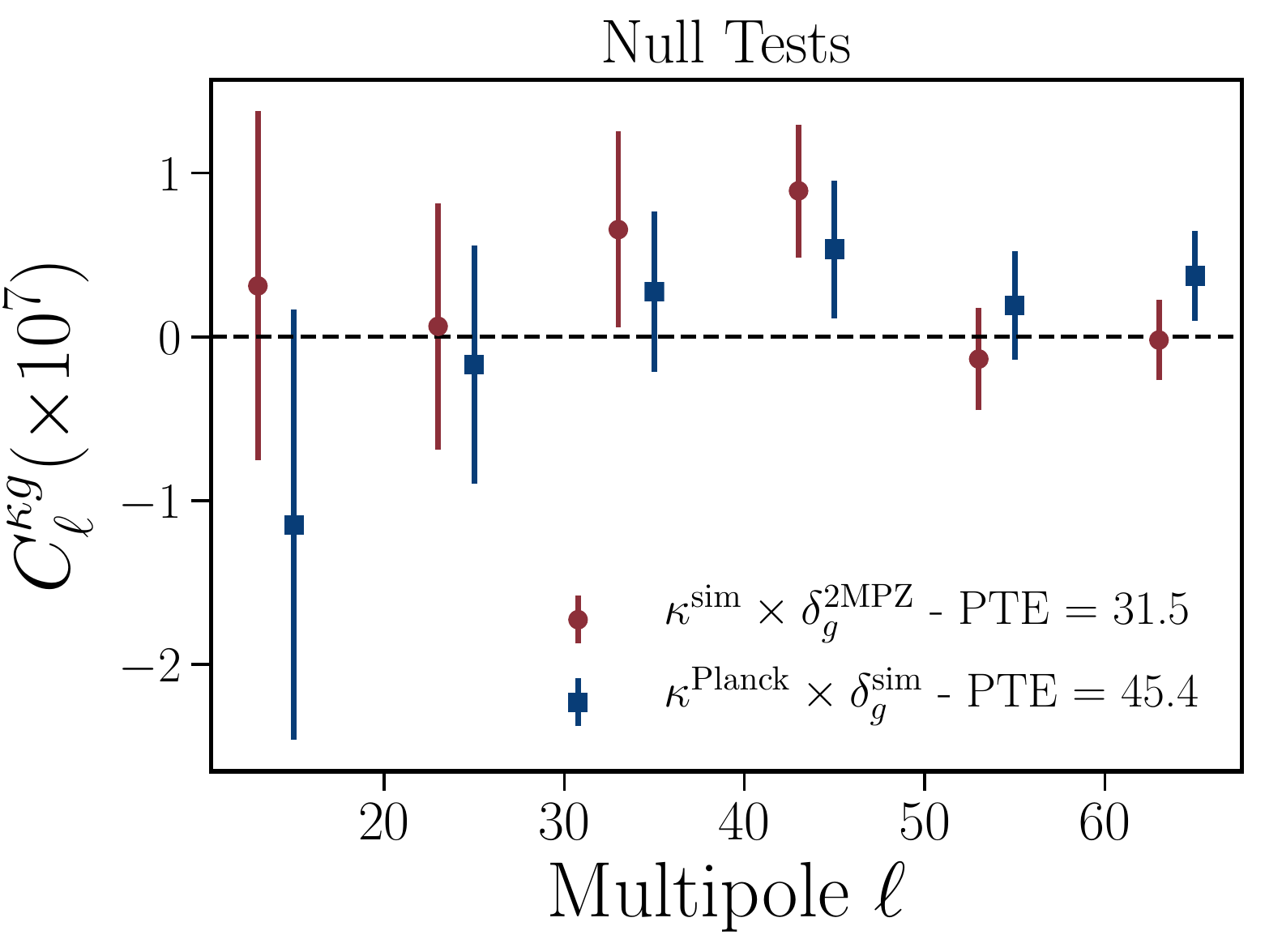}
    \caption{Null test results. Mean cross-power spectrum $C_{\ell}^{\kappa g}$ between the 2MPZ galaxy density map and the 100 simulated \textit{Planck} CMB lensing maps (red circles) and between the 100 mock galaxy maps and the observed \textit{Planck} lensing map (blue squares). Error bars are given by the scatter in the 100 cross-spectra divided by $\sqrt{100}$.}
    \label{fig:null_test_planck}
\end{figure}
We start by showing in Fig.~\ref{fig:null_test_planck} the mean cross-power spectrum between the true 2MPZ galaxy map and the set of 100 CMB lensing simulations released by the \textit{Planck} team (red circles).\footnote{\url{https://wiki.cosmos.esa.int/planckpla2015/index.php/Simulation\_data\#Lensing\_Simulations}} While these realizations capture the full complexity of the CMB lensing reconstruction analysis \citep{planck_lens_2015}, they do not contain any cosmological signal correlated with the spatial distribution of the 2MPZ catalogue. Hence, if the power spectrum extraction pipeline does not introduce any spurious correlation, the average is expected to be zero. In fact, for the scales in the range $10 \le \ell \le 70$, we find $\chi^2/\nu \sim 7.1/6$, corresponding to a probability-to-exceed (PTE) of about $31.5$ (assuming Gaussian random deviates). 
In Fig.~\ref{fig:null_test_planck}, we also show the measured mean cross-correlation signal between the observed \textit{Planck} CMB lensing map and a set of 100 galaxy mocks (blue squares). The  average cross-spectrum is consistent with null in this case too. Specifically, we obtain $\chi^2/\nu \sim 5.7/6$, corresponding to a PTE of $\sim 45.4$.
We then conclude that our extraction pipeline does not bias the observed power spectra.

\begin{figure*}
	\includegraphics[width=\textwidth]{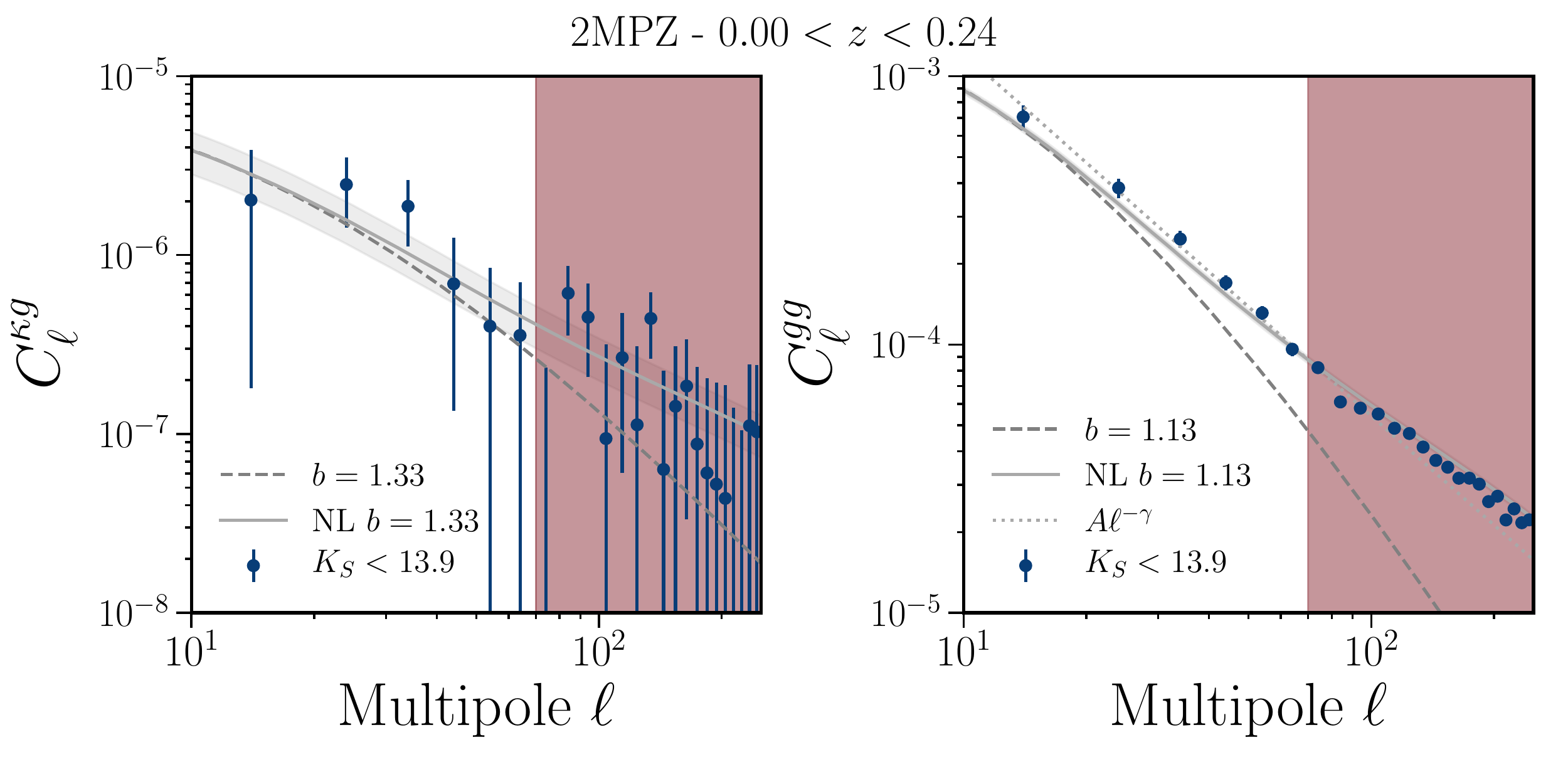}
    \caption{Extracted CMB lensing-galaxy cross- (\textit{left panel})  and galaxy auto-power spectrum (\textit{right panel}) for our baseline sample. In each panel, the recovered band powers are shown as blue circles (error bars given by Eq.~\ref{eqn:errors}), while the theoretical predictions including (excluding) non-linear corrections are shown as gray solid (dashed) lines. Theoretical lines are computed assuming the best-fit galaxy bias inferred from the cross- or auto-power spectrum respectively. The shaded gray areas denote the 1$\sigma$ region around the best-fit theory, while the shaded red regions highlight the range of multipoles discarded from the bias-fitting procedure. The dotted grey line in the right panel shows the best-fit power-law to the angular auto-power spectrum.}
    \label{fig:spectra}
\end{figure*}

\section{Results}
\label{sec:results}
We present here the results of the power spectrum analysis, as well as the constraints on the growth factor.

\subsection{Power spectra}
The extracted cross-power spectrum between the \textit{Planck} CMB lensing map and the 2MPZ catalogue is shown in Fig.~\ref{fig:spectra} (left panel), together with the galaxy auto-power spectrum (right panel). A clear cross-correlation signal can be seen, even though the CMB lensing kernel and 2MPZ redshift distribution are not perfectly matched (see Fig.~\ref{fig:2MPZ_dNdz}).

In Fig.~\ref{fig:spectra}, we show the best-fit curves and their $1\sigma$ uncertainties as the solid grey lines and contours respectively. A popular parametrization of the galaxy auto-power spectrum is the power-law $C_{\ell}=A\ell^{-\gamma}$. We have checked that the auto-power spectrum shape is well approximated by a power-law function with $A=0.027 \pm 0.007$ and $\gamma=1.35 \pm 0.06$. This is  broadly consistent with \citet{Balaguera2017} findings.

As a consistency check, we individually estimate the best-fit galaxy bias by comparing the observed $\hat{C}_{\ell}^{\kappa g}$ and $\hat{C}_{\ell}^{gg}$ to the theoretical predictions in the multipole range $10 \le \ell \le 70$.\footnote{These angular scales correspond
to physical separations of about $15 \lesssim r \lesssim 100$ Mpc at the catalog median redshift $z_{\rm med}\approx 0.08$.} 
To this end, we assume Gaussian likelihoods as $-2\ln\mathcal{L}(\mathbf{d}|\bm{\theta}) \propto \chi^2$, where $\chi^2 = [\mathbf{d}-\mathbf{t}(\bm{\theta})]^T\text{C}^{-1}[\mathbf{d}-\mathbf{t}(\bm{\theta})]$, $\mathbf{d}$ is the data vector (measured band powers), $\mathbf{t}(\bm{\theta})$ is the theory vector predicted by model parameters $\bm{\theta}$, and C$^{-1}$ is the covariance matrix. The posterior space $p(\bm{\theta}|\mathbf{d}) \propto \mathcal{L}(\mathbf{d}|\bm{\theta})  \pi(\bm{\theta})$ is then sampled via a Markov chain Monte Carlo (MCMC) method implemented in the public \texttt{emcee} code \citep{emcee}. The covariance matrix is assumed to be diagonal with elements given by Eq.~\ref{eqn:errors}, while priors $\pi(\bm{\theta})$ are taken to be flat. The cross-power spectrum analysis reveals a galaxy bias of $b_{\kappa g}= \bkg$ and a $\chi^2=2.3$ for $\nu =6-1=5$ degrees-of-freedom (DOF), corresponding to a PTE of $\sim 80\%$, while for the auto-power spectrum we find $b_{gg}=\bgg$ and a $\chi^2=5.2$ for $\nu=5$, translating to a PTE of approximately 39\%. Both these galaxy biases are consistent with each other and in good agreement with those found by \citet{Stolzner2017,Balaguera2017}.
For both the auto- and cross-correlation, we can estimate the significance of the detection by calculating $S/N = \sqrt{\chi^2_{\rm null}-\chi^2_{\rm bf}}$, where $\chi^2_{\rm null}=\chi^2(b=0)$ and $\chi^2_{\rm bf}$ is the $\chi^2$ evaluated at the best-fit. We find $S/N\approx 3.7$ and $\approx 36$ for the cross- and auto-power spectrum respectively, in good agreement with the values estimated in Sec.~\ref{sec:methods} albeit slightly lower in the case of cross-correlation. This might be due to an overestimation of the cross-spectrum uncertainties based on Eq.~\ref{eqn:errors} or to a simple statistical fluke.

\subsection{Constraints on the growth history}
\label{sec:dg_results}
\begin{figure}
	\includegraphics[width=\columnwidth]{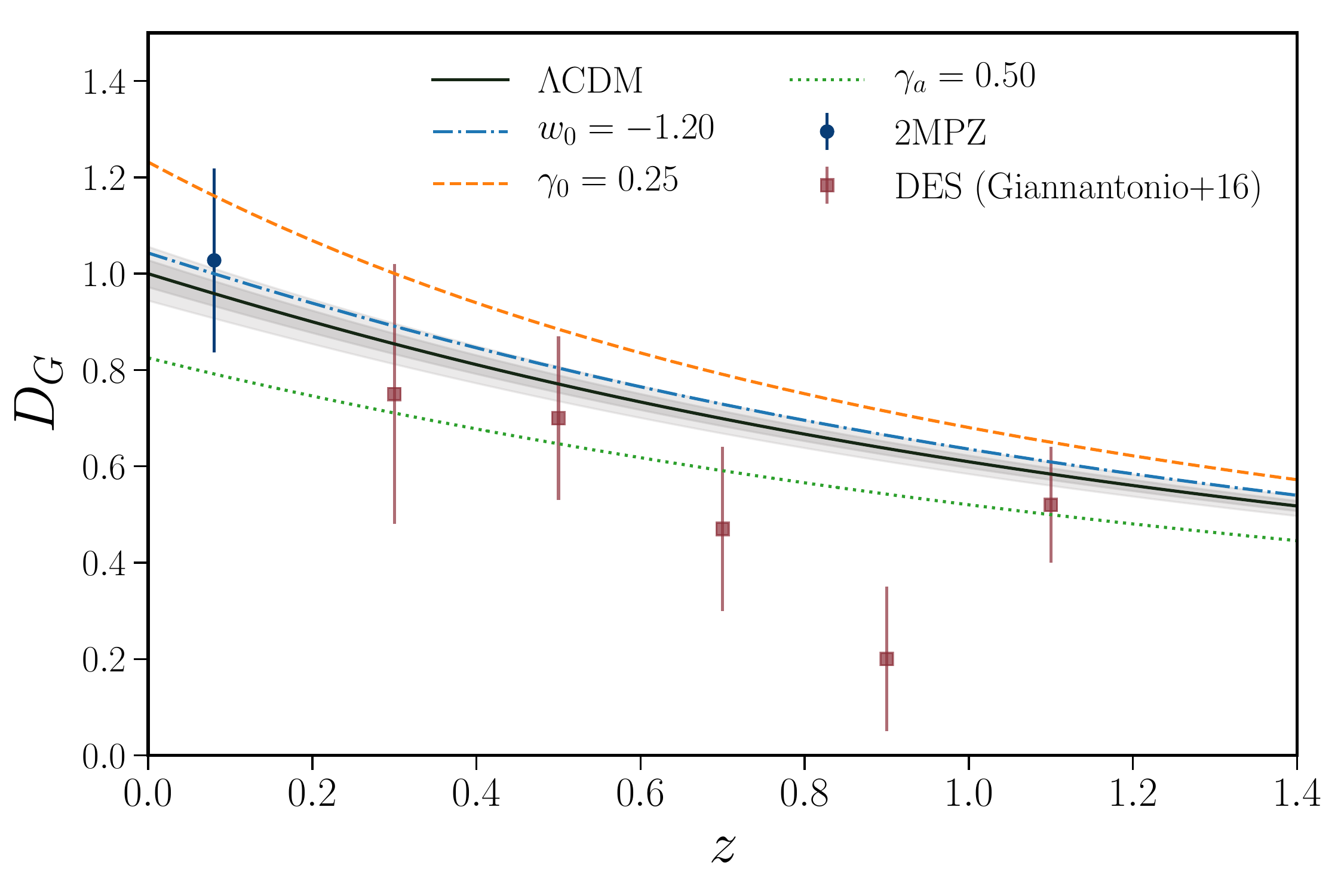}
    \caption{Linear growth factor estimated with the $D_G$ statistics applied to \textit{Planck} CMB lensing and 2MPZ data (blue circle). The solid blue line shows the theoretical growth function expected for the fiducial cosmology, while the light red square points display the tomographic $D_G$ measurement from DES \citep{Giannantonio2016}. The colored lines show the theoretical growth function expected in cosmologies different from $\Lambda$CDM. The dark and light grey bands represent the $1\sigma$ and $2\sigma$ scatter for 5000 cosmologies randomly drawn from the \textit{Planck} chains respectively.}
    \label{fig:DG_vs_z}
\end{figure}

After extracting the power spectra, we apply the $D_G$ estimator to the \textit{Planck} and 2MPZ datasets and obtain $\hat{D}_G = \dgvalue$ as shown in Fig.~\ref{fig:DG_vs_z}. The error bars are estimated in a Monte Carlo approach. To this end, we generate $N_{\rm sims} = 500$ correlated Gaussian realizations of the CMB convergence and galaxy fields with statistical properties that match the observed data (a thorough description of these steps can be found in \citet{Bianchini2015}). Then, we use the extracted auto- and cross-spectra from the simulation ensemble to obtain $N_{\rm sims}$ estimates of $D_{G}$. We quote the scatter across these estimates as our $1\sigma$ uncertainty on the measured $\hat{D}_G$. 

Following \citet{Giannantonio2016}, we assume the cosmic growth history shape $D_G(z)$ to be fixed at high redshift by the fiducial cosmology and then fit for its amplitude $A_D$ as $\hat{D}_G(z_{\rm med}) = A_D D_G^{\rm th}(z_{\rm med})$. The result of the fit is $A_D = 1.06 \pm 0.20$, in good agreement with $\Lambda$CDM. Note that the magnitude of the uncertainties $\Delta A_D$ is comparable to what found by \citet{Giannantonio2016} for the DES sample, although their result hints to a lower amplitude value, being $1.7\sigma$ away from $A_D=1$. We also stress that the two analyses are complementary in their redshift ranges: 
$0 \le z \le 0.24$ in this work versus  $0.2 \le z \le 1.2$ for the DES analysis. 

Of course, the expected signal does depend on cosmology, specifically the parameter combination, $\sigma_8\Omega_m H_0^2$. 
The high value observed for $D_G$ could be explained by increasing the $\sigma_8\Omega_m H_0^2$ combination by about $\approx 6\%$ over the fiducial cosmology. 
The dark and light gray regions in Fig.~\ref{fig:DG_vs_z} show the 1\,$\sigma$ and 2\,$\sigma$ variations in the predicted $D_G$ across a \textit{Planck} chain. 
Specifically,  we randomly draw 5000 points from \textit{Planck} chains and calculate the linear growth functions $D(z)$ for each of the 5000 cosmologies. 
We normalize the curve for model $i$ by multiplying by $(\sigma_8\Omega_m H_0^2)_i/(\sigma_8\Omega_m H_0^2)_{\rm fid.}$  \citep{Giannantonio2016}.
The uncertainty in cosmological parameters induces a scatter of approximately $\sigma_{D_G}\approx 0.03$. 

\begin{figure}
	\includegraphics[width=\columnwidth]{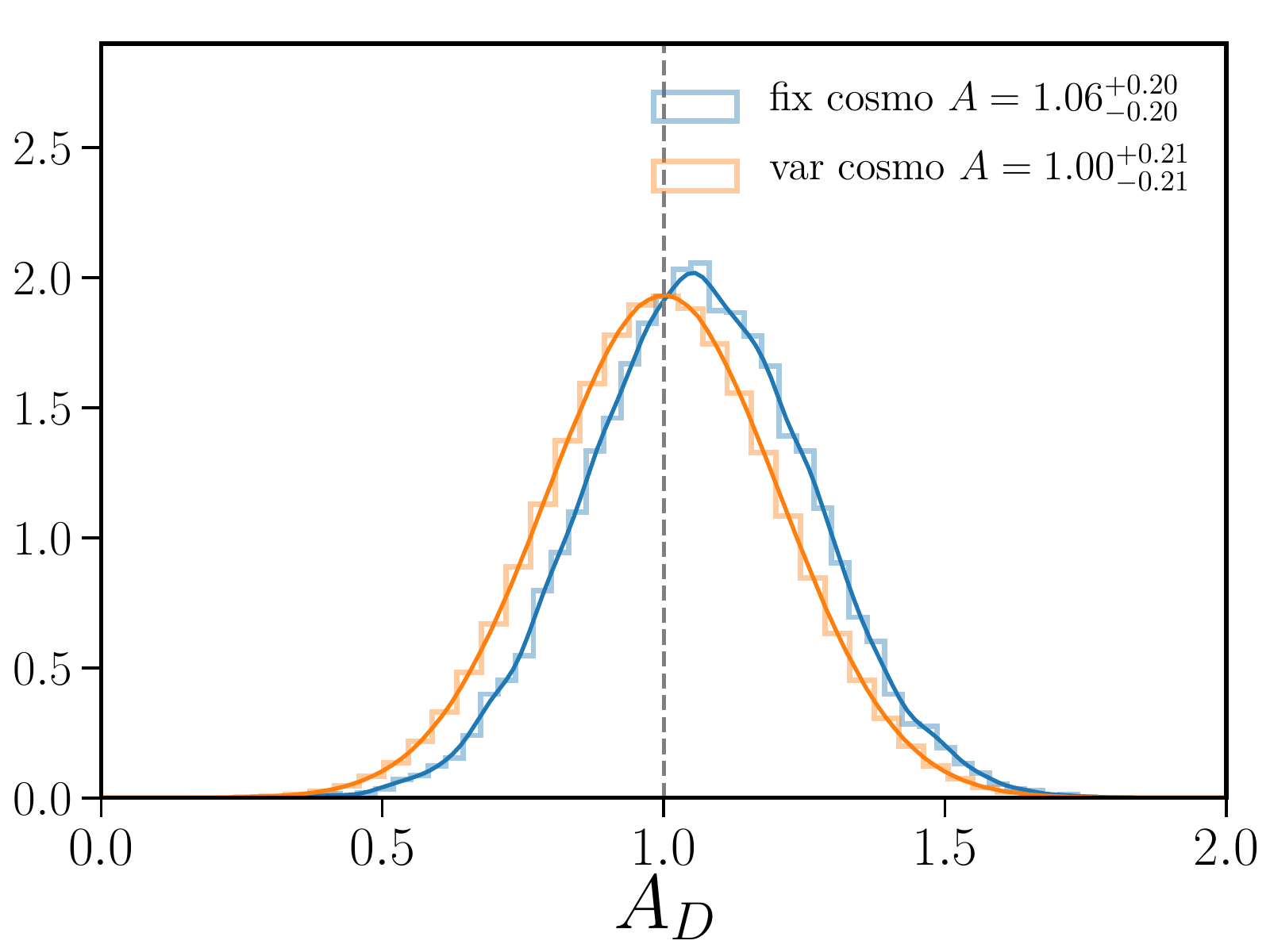}
    \caption{Observed growth factor amplitude $A_D$ is very consistent with $\Lambda$CDM, as can be seen from its posterior distribution when cosmological parameters are fixed (blue lines) or varying (orange lines). The smooth lines represent a kernel density estimate of the underlying histograms.}
    \label{fig:A_chains}
\end{figure}

While the scatter is small compared to the overall uncertainty, we choose to include the cosmological uncertainty into the $A_D$ estimate. 
We do this by stacking the $A_D$ likelihoods across 5000 randomly selected points, $\bm{\theta}_{\rm cosmo}^i$,  from the stationary \textit{Planck} chains.  
For each parameter set, we calculate the likelihood 
\begin{equation}
-2 \ln\mathcal{L}(\mathbf{d}|A_D,\bm{\theta}_{\rm cosmo}) \propto \frac{[\hat{D}_G(\bm{\theta}_{\rm cosmo})- A_D D_G^{\rm th}(\bm{\theta}_{\rm cosmo})]^2}{\Delta \hat{D}_G^2},
\end{equation}
at each step $i$. Then, we marginalize over the cosmological parameters by stacking the posteriors evaluated at each step. The resulting posterior distribution for the growth factor amplitude $A_D$ is shown in Fig.~\ref{fig:A_chains}. When we allow the underlying cosmological parameters to vary, we find $A_D=\advalue$, in excellent agreement with the $\Lambda$CDM expectations and in line with the constraint for the fiducial cosmology.

One of the main goals of structure probes like this $D_G$ measurement is to inform us about dark energy and modified gravity models. 
In Fig.~\ref{fig:DG_vs_z}, we explore the parameter space associated to more exotic scenarios in which we allow for departures from the standard $\Lambda$CDM. For the sake of simplicity, we consider modifications of the standard model where there is no induced scale-dependent growth, such as the Linder $\gamma$ parametrization \citep{Linder2005} of the growth rate $f(z)=[\Omega_{\rm m}(z)]^{\gamma}$, with  $\gamma(z)=\gamma_0+\gamma_a z/(1+z)$, or the dynamical dark energy model with time-dependent equation of state as $w(z)=w_0+w_a z/(1+z)$ \citep{cpl}. Note that to highlight the effect of changing the single dark energy/modified gravity parameters in Fig.~\ref{fig:DG_vs_z}, we fix $\gamma_0$ ($w_0$) to its standard value $\gamma_0=0.55$ ($w_0=-1$) when varying $\gamma_a$ ($w_a$) and vice versa. 
As done in the previous section, these predictions have been normalized such that $D_G^i$ matches the $\Lambda$CDM one at high-$z$ and apply the $(\sigma_8\Omega_m H_0^2)_i/(\sigma_8\Omega_m H_0^2)_{\rm Planck}$ rescaling \citep{Giannantonio2016}. 
Although the data is not yet sufficiently sensitive to allow for an interesting test of models, the result seems to favor values of the dark energy equation of state $w_0 < -1$, or $\gamma_0$ values smaller than the predicted value in GR, $\gamma_0\approx 0.55$. 
As we will show in \S\ref{sec:forecast}, future experiments will be able to discriminate between these models.
Note also that combining measurements of $f(z)\sigma_8(z)$ from RSD with that of the linear growth factor $D(z)$  at a given redshift can potentially break the degeneracy between the cosmic matter density $\Omega_m$ and the growth index $\gamma$.

\subsection{Robustness against analysis choices}
\label{sec:robustness}
\begin{figure}
	\includegraphics[width=\columnwidth]{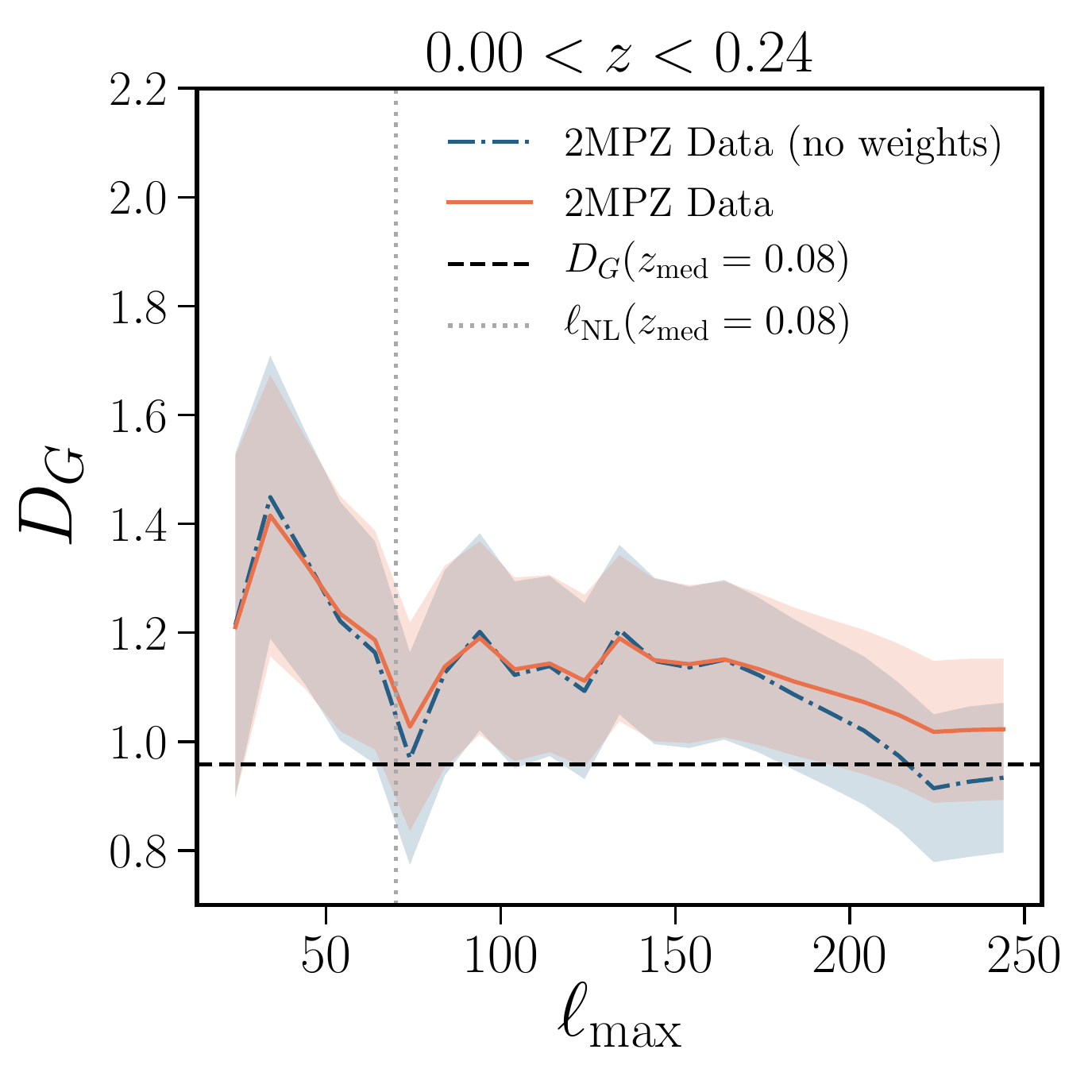}
    \caption{Results are relatively insensitive to exact choice of $\ell_{\rm max}$. Here, we show the inferred growth factor as function of the maximum multipole used in the $D_G$ estimator. The solid red (dash-dotted blue) line denotes the result when the inverse-variance weighting is (not) applied to $D_G$, while the shaded areas represent the scatter from simulations. The vertical dotted line indicates the angular scale subtended by the modes that are entering the non-linear regime at $z\simeq 0.08$.}
    \label{fig:DG_lmax}
\end{figure}

We now turn to investigating the impact of different analysis choices on the results.
The first choice we consider is the range of  angular scales used in the $D_G$ estimator. 
Effectively, the question is to whether extending the range into the non-linear regime leads to a shift in the observed $D_G$ value. 
To zeroth order, we would expect non-linear structure growth to cancel out for similar reasons as the bias cancellation. 
However if non-linear structure growth is biasing the result, we should see a monotonic shift in $D_G$ as the maximum multipole is increased. 
In Fig.~\ref{fig:DG_lmax}, we show the estimated $\hat{D}_G$ value as function of the maximum multipole $\ell_{\rm max}$ included in Eq.~\ref{eqn:dg}. The shaded regions reflect the 1\,$\sigma$ scatter in the growth factor estimates across the $N_{\rm sims}$ simulations presented in Sec.~\ref{sec:dg_results}. 
The dotted vertical line reflects the angular multipole of modes that are entering the non-linear regime at $z=0.08$. 
There is no significant shift in the $D_G$ value with $\ell_{\rm max}$, and we conclude that is unlikely that the inferred amplitude of the growth factor is affected by the improper inclusion of non-linear scales.

\begin{figure}
	\includegraphics[width=\columnwidth]{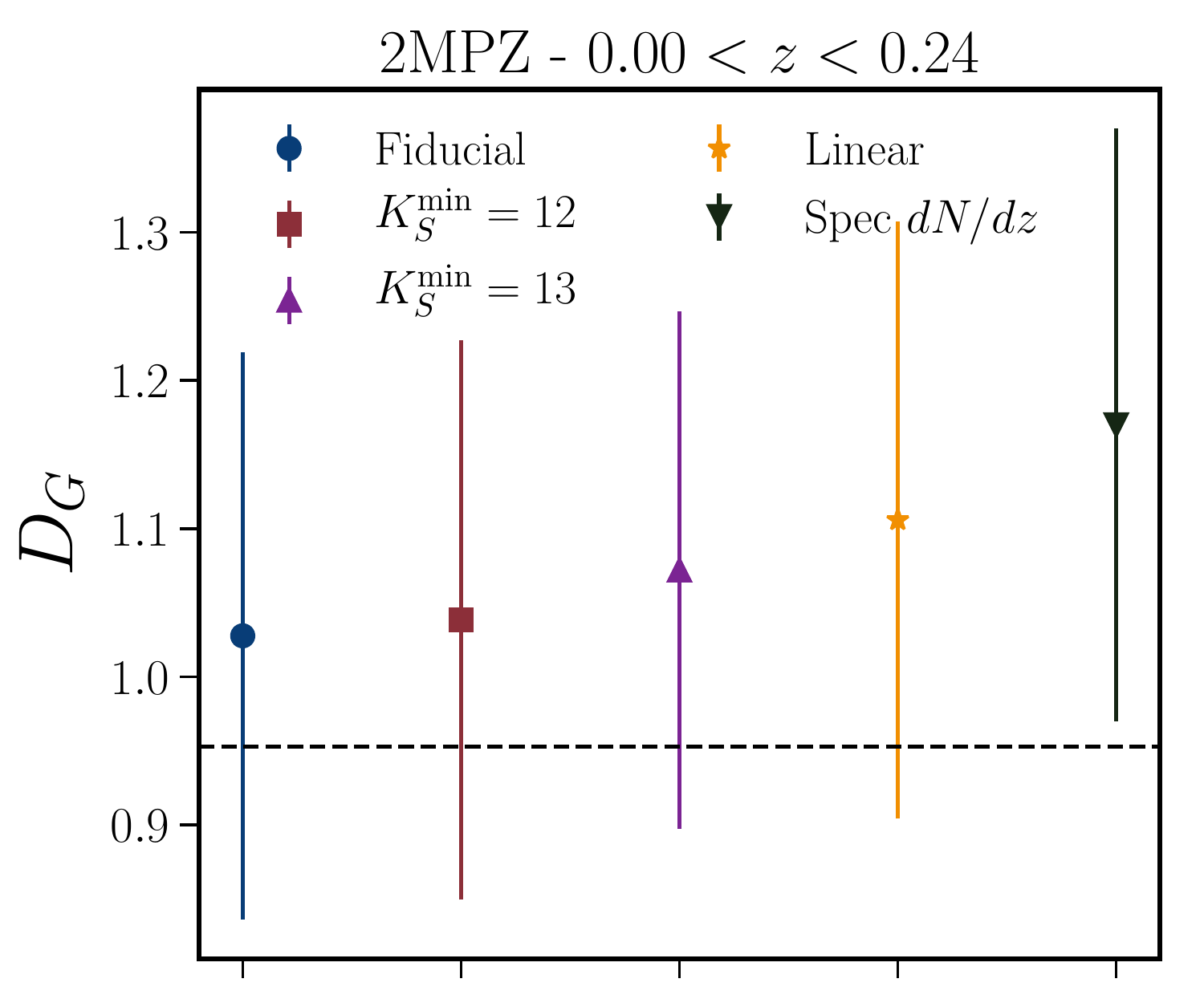}
    \caption{Effect of different analysis choices on the estimated growth factor. The (blue) circle denotes the fiducial $D_G$ value (non-linear corrections and no magnitude cuts applied), while the (red) square and the (violet) upward triangle represent the results when $K_S^{\rm min}=12,13$ is assumed respectively. The (yellow) star denotes the results without the inclusion of the non-linear corrections in the theoretical spectra, while the (black) downward triangle shows the results when the spectroscopic $dN/dz$ is assumed to represent the whole 2MPZ sample. $D_G$ is assumed to be averaged over $10 \le \ell \le 70$ in all cases.}
    \label{fig:DG_systematics}
\end{figure}

In the baseline analysis, we compute the theoretical power spectra including  non-linear corrections from \texttt{Halofit}. 
We test the impact of this decision in Fig.~\ref{fig:DG_systematics}, where we show the $D_G$ values estimated with and without the inclusion of such corrections. 
As can be seen, the impact is rather mild, with the $D_G$ value in the linear case drifting towards higher values but still within $\sim 0.4\sigma$  of the baseline result. 
This can be understood as follows. The non-linear impact at low redshift is more pronounced for the galaxy auto-power spectrum rather than the cross-spectrum, meaning that $\sqrt{\slashed{C}_{\ell}^{gg,\text{lin}}/\slashed{C}_{\ell}^{gg, \text{nl}}} > (\slashed{C}_{\ell}^{\kappa g,\text{lin}}/\slashed{C}_{\ell}^{\kappa g, \text{nl}})$. 
Then, it follows that $D_G^{\rm lin} > D_G^{\rm nl}$.

In Fig.~\ref{fig:DG_lmax}, we also show the impact of using uniform weighting (dot-dash blue) instead of the baseline inverse-variance weighting (solid orange) of modes entering the $D_G$ estimator. 
The inverse-variance weighting de-weights the higher noise high angular modes, whereas these noisy modes can pull the estimate around in the uniform weight case. 
As a result, the uniform weighting case is somewhat more dependent on the choice of $\ell_{\rm max}$. 
When the weights are applied, the values are more stable and almost independent on $\ell_{\rm max}$. 

We also test the dependence of the results on the limiting lower magnitude in the $K$-band. By raising the threshold $K_S^{\rm min}$ we progressively select intrinsically brighter, therefore more biased, objects at higher redshifts. We show the results in Fig.~\ref{fig:DG_systematics} and conclude that results are stable against magnitude cuts.
%By raising the threshold $K_S^{\rm min}$ we progressively select fainter and less biased objects. We show the results in Fig.~\ref{fig:DG_systematics} and conclude that results are stable against magnitude cuts.
%

The knowledge of how galaxies are statistically distributed as function of redshift is a key ingredient of the present analysis as it allows us to predict the theoretical power spectra entering Eq.~\ref{eqn:dg}. Any mismatch between the true and assumed redshift distribution could potentially bias the inferred value of the growth factor. The first test that we conduct is to assume the redshift distribution of the spectroscopic 2MPZ subsample (orange histogram in Fig.~\ref{fig:2MPZ_dNdz}) to be representative of the whole sample. By doing so, we are effectively testing the robustness of the analysis with respect to systematic shifts in the assumed galaxy redshift distribution, as the spectroscopic $dN/dz$ peaks towards lower redshifts than the fiducial one because most of the sources with spec-$z$ comes from the shallow subsamples of 2MPZ, 2MRS and 6dFGS (see Fig.~\ref{fig:2MPZ_dNdz}). In this case, we find best-fit linear galaxy biases of $b_{\kappa g} = 1.44 \pm 0.40$ and $b_{gg} = 1.06 \pm 0.03$ for the cross- and auto-power spectrum respectively. While the bias constraint from $C_{\ell}^{\kappa g}$ is almost unaffected by changes in the $dN/dz$, the best-fit galaxy bias from $C_{\ell}^{gg}$ decreases by $\approx 9\%$ to compensate the shift in the redshift distribution. In turn, this translates into a growth factor of about $\hat{D}_G = 1.17 \pm 0.20$, approximately 14\% higher than our baseline value but still within $0.7\sigma$ (see Fig.~\ref{fig:DG_systematics}). We stress that the result of this test does not imply the presence of any systematic effect since we have purposely input an erroneous redshift distribution (the spec $dN/dz$) while photo-$z$s are known to be virtually unbiased. We then test for the smearing effect of photo-$z$ error by performing the analysis considering the full photometric $dN/dz$ and taking $\sigma_z = 0$ and 0.03. For the two cases respectively we find $\hat{D}_G = 1.03 \pm 0.19$ and $\hat{D}_G = 1.02 \pm 0.19$, meaning that a broadening or narrowing of the $dN/dz$ distribution has a negligible impact on the result.

As a final remark, we note that care has to be taken when interpreting results based on the $D_G$ as well as the $E_G$ estimators. The main reason is that there is a mismatch between the lensing and clustering kernels, meaning that these two measurements probe structures at different effective redshifts. A further complication to this picture is represented by the scale and redshift dependence of the clustering bias. In order to account for these effects, correction factors have been devised in the context of $E_G$ measurements \citep{Reyes2010}. For example, \citet{Pullen2016} found that corrections to the $E_G$ estimator are $\lesssim 6\%$ out to scales of about $\ell \sim 400$ in the case of the CMASS-\textit{Planck} lensing cross-correlation. Given the 20\% statistical error characterizing the $D_G$ measurement presented in this work, we neglect these systematics corrections but caution the reader that they could affect the analysis of forthcoming CMB and LSS datasets, hence requiring a careful modeling using simulations.

\subsection{Forecasts for future CMB and LSS surveys}
\label{sec:forecast}
\begin{figure*}
	\includegraphics[width=\textwidth]{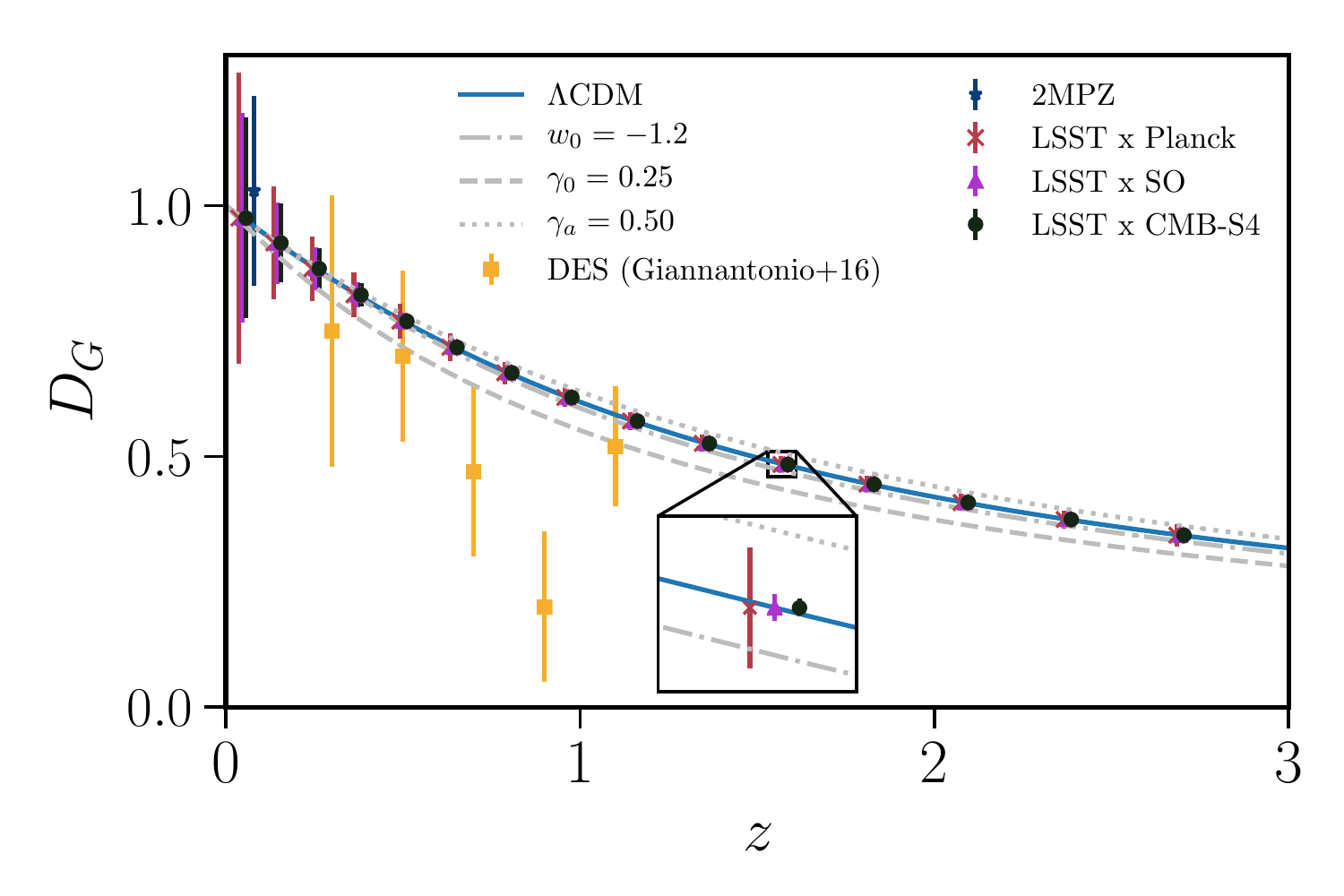}
    \caption{$D_G$ forecasts for the LSST photometric galaxy survey cross-correlated with \textit{Planck} (light red crosses), Simons Observatory (purple triangles), and CMB-S4 lensing maps (black circles). Points are offset by $\Delta z = \pm 0.01$ for visualization purposes. The current 2MPZ measurement is shown as a dark blue star, while the yellow square points represent the DES measurement from \citet{Giannantonio2016}. The solid blue line represents the linear growth factor in the standard $\Lambda$CDM scenario, while the grey lines show $D(z)$ for different dark energy/modified gravity models. Note that in this case we have not applied the $\sigma_8\Omega_{\rm m} H_0^2$ rescaling as in Fig~\ref{fig:DG_vs_z}.}
    \label{fig:D_G_forecast}
\end{figure*}
In this subsection, we look to the future and predict the capability of future LSS and CMB surveys to measure $D_G$. For a fixed area of overlapping sky, the key factors that determine the overall $S/N$ are the lensing reconstruction noise and the galaxy number density.

On the optical side, we consider a Large Synoptic Survey Telescope (LSST)-like photometric redshift survey \citep{lsst}. LSST is expected to image few billions of galaxies over a large range of redshifts, optimally overlapping with the CMB lensing kernel. We assume the LSST "gold" sample, defined by a magnitude limit of 25.3 in the $i$ band, to be our representative galaxy sample. This is expected to include $\sim 40$ galaxies/arcmin$^{2}$ with the photometric redshift distribution being well approximated by the following functional form
\begin{equation}
\frac{dN}{dz} \propto \left( \frac{z}{z_0}\right)^2 \exp\left(-\frac{z}{z_0}\right).
\end{equation}
Here, $z_0 = 0.0417i - 0.744$, resulting in a median of about $0.8$ given our magnitude cut choice. We model the linear galaxy bias evolution as $b(z) = 1 + 0.84z$, accordingly to the LSST science book \citep{lsst}. The photo-$z$ errors requirement for the LSST gold sample is $\sigma_z/(1+z) < 0.05$ with a goal of $0.02$. Here we take a conservative approach and assume an intermediate value of $\sigma_z(z)=0.03(1+z)$ and split the main galaxy sample in photo-$z$ bins of width given by $3 \times \sigma_z(\bar{z})$, where $\bar{z}$ is the bin centre \citep{Alonso2017}. This choice results in 15 redshift bins.

On the CMB side, we consider two upcoming ground-based  surveys, the Simons Observatory\footnote{\url{https://simonsobservatory.org}} (SO) and CMB Stage-4 \citep[CMB-S4]{cmbs4}. These surveys will provide multi-frequency measurements of the microwave sky and deliver high $S/N$ maps of the CMB lensing convergence. In order to estimate the CMB lensing noise curves, we consider a $\theta_{\rm FWHM}= [ 1.4, 2]$ arcmin beam and $\Delta_T = [6, 1]\,\mu$K-arcmin noise level for SO and CMB-S4 respectively. We also assume the lensing reconstruction to be performed with  CMB modes from about $\ell_{\rm min} = 30$ up to $\ell_{\rm max}^{E,B}=5000$ for polarization and up to $\ell_{\rm max}^T=3000$  for temperature, to reflect the impact of foregrounds in the intensity maps.

In our forecasting setup, we consider the LSS and CMB surveys to overlap over an area of about 18,000 deg$^2$. The expected uncertainties are calculated by assuming that the auto- and cross-power spectra will be measured from $\ell_{\rm min}=10$, reflecting the difficulty to recover the largest scales from ground observations, up to a redshift dependent cutoff multipole given by $\ell_{\rm max}(z) = k_{\rm NL}(z) \chi(z)$, in order to avoid the inclusion of non-linear scales. This cutoff scale goes from $\ell_{\rm max} \approx 30$ at low redshift, up to more than 3000 for higher redshift. We show the forecasted $D_G$, along with the current measurements, in Fig.~\ref{fig:D_G_forecast}. To give a rough estimate of how the sensitivity to $D_G$ varies across the experimental landscape, we calculate the total $S/N$ integrated over angular scales and redshift bins $z_i$ as 
\begin{equation}
S/N = \sqrt{\sum_{z_i} \left( \frac{D_G(z_i)}{\Delta D_G(z_i)}\right)^2  }. 
\end{equation}
We can also predict at what significance level a certain datasets combination can differentiate between standard $\Lambda$CDM and a given alternative model. To this end, we calculate 
\begin{equation}
\chi^2 = \sum_{z_i} \left( \frac{D_G^{\rm DE/MG}(z_i) - D_G(z_i)}{\Delta D_G(z_i)} \right)^2, 
\end{equation}
where $D_G$ and $D_G^{\rm DE/MG}$ are the growth factor calculated for $\Lambda$CDM and a certain dark energy/modified gravity model respectively. Then, we can quote $\sqrt{\chi^2}$ as the significance of the discrimination between two scenarios
\citep{Pullen2015}. As can be seen in Tab.~\ref{tab:D_G_forecast}, LSST high galaxy number density will allow for high $S/N$ measurements of $D_G$, making possible the discrimination between different exotic models at high significance. Specifically, the lower lensing reconstruction noise that characterizes the forthcoming CMB surveys will improve the overall $S/N$ by a factor 3.4 and 5 with respect to \textit{Planck} for SO and CMB-S4, respectively.

\begin{table*}
\centering
\caption{Total $S/N$ of the $D_G$ measurement for the LSST photo-$z$ survey combined with current and upcoming CMB surveys. We also report the significance of discrimination between $\Lambda$CDM and a CPL model with $w_0$=-1.2, and two Linder models with $\gamma_0 = 0.25$ and one with $\gamma_a=0.5$.}
\label{tab:D_G_forecast}
\begin{tabular}{c|cccc}
Survey        & $S/N$ & $\sqrt{\chi^2} \quad [w=-1.2] $ & $\sqrt{\chi^2} \quad [\gamma_0=0.25] $ & $\sqrt{\chi^2} \quad [\gamma_a=0.5] $ \\ \hline
LSST x Planck & 92    & 2.1                             & 8.7                                    & 3.5                                   \\
LSST x SO     & 312   & 8.2                             & 31.3                                     & 13.4                                  \\
LSST x CMB-S4 & 468   & 13.1                            & 48.3                                     & 21.3                                 
\end{tabular}
\end{table*}
\newpage
\section{Conclusions}
\label{sec:conclusions}
In this paper, we have performed a new consistency test of the $\Lambda$CDM model by measuring the linear growth factor at $z\simeq 0.08$. To this end, we have combined the cross-power spectrum between the \textit{Planck} CMB lensing and the 2MPZ galaxies with the galaxy auto-power spectrum into the (bias-independent) $D_G$ estimator introduced by \citet{Giannantonio2016}. Our result is in agreement with the $\Lambda$CDM scenario, suggesting an observed growth factor of about $\hat{D}_G = \dgvalue$, corresponding to a structure growth amplitude of $A_D = \advalue$ . 
This work extends $D_G$ measurements to the local universe ($z\lesssim 0.2$) and over much larger sky fractions. 

The combination of CMB lensing and clustering measurements offers an exciting avenue to test the $\Lambda$CDM model and its extensions on cosmological scales \citep{Schmittfull2017}.  
Although this measurement is not yet sensitive enough to rule out deviations from a $\Lambda$CDM growth history, this work represents a preliminary step towards the challenges posed by the analysis of the new generation of LSS and CMB datasets such as LSST \citep{lsst}, Euclid \citep{euclid}, WFIRST, SPT-3G \citep{spt3g}, Advanced ACTPol \citep{advact}, Simons Array \citep{simonsarray}, Simons Observatory, and CMB-S4 \citep{cmbs4}. At the same time, while statistical errors keep shrinking thanks to the augmented experiments sensitivity, the need for an accurate theoretical modeling becomes indispensable \citep{modi2017}. As we have forecasted, LSST combined with SO and CMB-S4 will provide sub-percent measurements of $D_G$ over a large range of redshifts and put tight constraints on dark energy/modified gravity models.


\acknowledgments
We are indebted to Chris Blake and Srinivasan Raghunathan for a careful reading of the draft and for enlightening discussions. We also thank Maciej Bilicki for providing valuable feedback on the manuscript. FB and CR acknowledge support from an Australian Research Council Future Fellowship (FT150100074). This research
has made use of data obtained from the SuperCOSMOS Science
Archive, prepared and hosted by the Wide Field Astronomy
Unit, Institute for Astronomy, University of Edinburgh,
which is funded by the UK Science and Technology
Facilities Council.
In this paper we made use of  \texttt{HEALPix} \citep{Gorski2005a}, \texttt{healpy}, and the \textit{Planck} Legacy Archive.


% \vspace{5mm}
% \facilities{HST(STIS), Swift(XRT and UVOT), AAVSO, CTIO:1.3m,


% \software{astropy \citep{2013A&A...558A..33A},  
%           Cloudy \citep{2013RMxAA..49..137F}, 
%           SExtractor \citep{1996A&AS..117..393B},
%           }


\appendix

\section{Inverse-variance weights}
\label{sec:weights}
The variance in the $D_G$ estimator can be calculated as
\begin{equation}
\Delta \hat{D}_G^2 = \frac{1}{N_L^2} \sum_L \hat{D}_{G,L}^2 \left[\left(\frac{\Delta \hat{C}_L^{\kappa g}}{\hat{C}_L^{\kappa g}} \right)^2 +  \frac{1}{4} \left(\frac{\Delta \hat{C}_L^{gg}}{\hat{C}_L^{gg}} \right)^2 \right],
\end{equation}
where $N_L$ is the number of bandpowers and the $D_G$ per each bandpower $L$ can be expressed as:
\begin{equation}
\hat{D}_{G,L} = \frac{\hat{C}_{L}^{\kappa g}}{\slashed{C}_{L}^{\kappa g}}\sqrt[]{\frac{\slashed{C}_{L}^{gg}}{\hat{C}_{L}^{gg}}}.
\end{equation}
Then, we can write the $D_G$ estimator as a weighted average across multipoles:
\begin{equation}
\hat{D}_G = \frac{\sum_{L}w_L \hat{D}_{G,L}}{\sum_L w_L},
\end{equation}
where the weights are given by
\begin{equation}
w_L^{-1} = \Delta \hat{D}_{G,L}^2 = \hat{D}_{G,L}^2 \left[\left(\frac{\Delta \hat{C}_L^{\kappa g}}{\hat{C}_L^{\kappa g}} \right)^2 +  \frac{1}{4} \left(\frac{\Delta \hat{C}_L^{gg}}{\hat{C}_L^{gg}} \right)^2 \right].
\end{equation}
%




\bibliography{references}

% \begin{thebibliography}{}

% \bibitem[Astropy Collaboration et al.(2013)]{2013A&A...558A..33A} Astropy Collaboration, Robitaille, T.~P., Tollerud, E.~J., et al.\ 2013, \aap, 558, A33 
% \bibitem[Bertin \& Arnouts(1996)]{1996A&AS..117..393B} Bertin, E., \& Arnouts, S.\ 1996, \aaps, 117, 393 
% \bibitem[Corrales(2015)]{2015ApJ...805...23C} Corrales, L.\ 2015, \apj, 805, 23
% \bibitem[Ferland et al.(2013)]{2013RMxAA..49..137F} Ferland, G.~J., Porter, R.~L., van Hoof, P.~A.~M., et al.\ 2013, \rmxaa, 49, 137
% \bibitem[Hanisch \& Biemesderfer(1989)]{1989BAAS...21..780H} Hanisch, R.~J., \& Biemesderfer, C.~D.\ 1989, \baas, 21, 780 
% \bibitem[Lamport(1994)]{lamport94} Lamport, L. 1994, LaTeX: A Document Preparation System, 2nd Edition (Boston, Addison-Wesley Professional)
% \bibitem[Schwarz et al.(2011)]{2011ApJS..197...31S} Schwarz, G.~J., Ness, J.-U., Osborne, J.~P., et al.\ 2011, \apjs, 197, 31  
% \bibitem[Vogt et al.(2014)]{2014ApJ...793..127V} Vogt, F.~P.~A., Dopita, M.~A., Kewley, L.~J., et al.\ 2014, \apj, 793, 127  

% \end{thebibliography}



\end{document}